\documentclass{article}
\usepackage{times}
\usepackage{amsfonts}
\usepackage{graphicx}
\begin{document}
\noindent
{\Large  ON THE HOLOGRAPHIC BOUND IN NEWTONIAN\\ COSMOLOGY}
\noindent

\vskip.5cm
\noindent
{\bf P. Fern\'andez de C\'ordoba}$^{a}$ and {\bf  J.M. Isidro}$^{b}$\\
Instituto Universitario de Matem\'atica Pura y Aplicada,\\ Universidad Polit\'ecnica de Valencia, Valencia 46022, Spain\\
${}^{a}${\tt pfernandez@mat.upv.es}, ${}^{b}${\tt joissan@mat.upv.es}\\

%\vskip.5cm
%\noindent
%\today
\vskip.5cm
\noindent
{\bf Abstract}  The holographic principle sets an upper bound on the total (Boltzmann) entropy content of the Universe at around $10^{123}k_B$ ($k_B$ being Boltzmann's constant). In this work we point out the existence of a remarkable duality between nonrelativistic quantum mechanics on the one hand, and Newtonian cosmology on the other. Specifically, nonrelativistic quantum mechanics has a quantum probability fluid that exactly mimics the behaviour of the cosmological fluid, the latter considered in the Newtonian approximation. One proves that the equations governing the cosmological fluid (the Euler equation and the continuity equation) become the very equations that govern the quantum probability fluid after applying the Madelung transformation to the Schroedinger wavefunction. Under the assumption that gravitational equipotential surfaces can be identified with isoentropic surfaces, this model allows for a simple computation of the gravitational entropy of a Newtonian Universe. 

In a first approximation we model the cosmological fluid as the quantum probability fluid of free Schroedinger waves. We find that this model Universe saturates the holographic bound. As a second approximation we include the Hubble expansion of the galaxies. The corresponding Schroedinger waves lead to a value of the entropy lying three orders of magnitude below the holographic bound. Current work on a fully relativistic extension of our present model can be expected to yield results in even better agreement with empirical estimates of the entropy of the Universe.

\section{Introduction}

There is a widespread belief that the continuum description of spacetime as provided by general relativity must necessarily break down at very short length scales and/or very high curvatures. A number of very different approaches to an eventual theory of quantum gravity have been presented in the literature; these candidate theories are too varied and too extensive to summarise here. Suffice it to say, though, that whatever the {\it atoms of spacetime}\/ may turn out to be, at the moment there exists a large body of well--established knowledge concerning the {\it thermodynamics of spacetime}\/. For recent advances in this direction, as well as more detailed bibliography, we refer the reader to the original articles \cite{PADDY1, PADDY2, PADDY3, PADDY4} as well as the review papers \cite{PHILO, MOUSTOS}. 

On the whole, the picture that emerges is that of a continuum description after some appropriate coarse graining of some underlying degrees of freedom (the atoms of spacetime mentioned above). Even if the precise nature of the latter is unknown yet, one can still make progress following a thermodynamical approach: one ignores large amounts of detailed knowledge (say, the precise motions followed by the atoms of a gas) while concentrating only on a few coarse--grained averages (say, the overall pressure exerted by the atoms of a gas on the container walls). This way of approaching the problem has come to be called {\it the emergent approach}\/.

In the emergent approach to spacetime presented in ref. \cite{VERLINDE}, gravity qualifies as an entropic force. Roughly speaking, this is the statement that we do not know the fundamental degrees of freedom underlying gravity, but their overall macroscopic effect is that of driving the system under consideration in the direction of increasing entropy. If gravitational forces are entropy gradients, then gravitational equipotential surfaces can be identified with isoentropic surfaces. This insight justifies identifying the gravitational potential and the entropy function (up to dimensional factors).

Recalling the arguments of ref. \cite{VERLINDE}, a classical point particle approaching a holographic screen causes the entropy of the latter to increase by one quantum $k_B$. We will replace the classical particle of ref. \cite{VERLINDE} with a density of particles representing the (baryonic and dark) matter contents of a hypothetical Newtonian Universe. This volume density will be identified with the squared modulus of a nonrelativistic wavefunction $\psi$ satisfying the Schroedinger equation. Let $U$ denote the gravitational potential. Once dimensions are corrected (using $\hbar$ and  $k_B$), the expectation value $\langle\psi\vert U\vert\psi\rangle$ becomes the quantum--mechanical analogue of the entropy increase caused by a classical particle approaching a holographic screen in ref. \cite{VERLINDE}. Therefore {\it the expectation value $\langle\psi\vert U\vert\psi\rangle$ becomes a measure of the gravitational entropy of the Universe when the matter  of the Universe is described by the wavefunction $\psi$}\/.  

The next question is to determine the Newtonian potential $U$ governing the Universe as a whole. Of course, even within the Newtonian approximation, $U$ necessarily appears as a very rough average. We can however find guidance in the Hubble expansion of the Universe \cite{HUBBLE, PERLMUTTER, RIESS}, which holds reasonably well over cosmological distances. This receding behaviour of the galaxies can be easily modelled  by a phenomenological potential, namely, an isotropic harmonic potential carrying a negative sign: 
\begin{equation}
U_{\rm Hubble}({\bf r})=-\frac{H_0^2}{2}{\bf r}^2. 
\label{potenzi}
\end{equation}
As the angular frequency  we take the current value of Hubble's constant $H_0$. (Thus $U_{\rm Hubble}$ has the dimensions of energy per unit mass, or velocity squared). The potential $U_{\rm Hubble}$ encodes the combined effect of the gravitational attraction, and of the repulsion caused by the dark energy on the matter content of the Universe (baryonic and dark matter).  We can therefore identify the Hubble potential $U_{\rm Hubble}$ of Eq. (\ref{potenzi}) with the gravitational potential $U$ in the previous paragraph. 

{}Following ref. \cite{BONDI}, let us briefly recall  why $U_{\rm Hubble}$ in fact combines a Newtonian gravitational attraction, plus a harmonic repulsion.\footnote{See Eq. (9.14 b) of ref. \cite{BONDI},  the right--hand side of which is the force that one would obtain by differentiation of our Eq. (\ref{potenzi}). The fact that ref. \cite{BONDI} defended the Steady State theory, the rival that lost against the Big Bang theory currently accepted, has no bearing on this discussion, as the Newtonian limit is the same.} In the Newtonian limit considered throughout in this paper, the gravitational attraction is computed by applying Gauss' law to a sphere filled with a homogeneous, isotropic density of matter. Then the gravitational field {\it within the sphere}\/ turns out to be proportional to the position vector, so the corresponding potential becomes a quadratic function of the position. Altogether, the total potential at any point within the cosmological fluid is the sum of two harmonic potentials; Hubble's constant $H_0$ is the frequency of this total harmonic potential. 

In this way the Newtonian space $\mathbb{R}^3$ is foliated by a continuous succession of concentric spheres with growing radii. Each one of these spheres qualifies as a gravitational equipotential surface. By what was said above, these surfaces are also isoentropic surfaces, the gradients thereto pointing in the direction of the gravitational force. The negative sign in Eq. (\ref{potenzi}) expresses the essential fact that this net force is repulsive instead of attractive. Already at the classical level, this potential possesses no state of least energy; a problem that resurfaces at the quantum level, as the inexistence of a stable vacuum state \cite{BROADBRIDGE}. What saves the day is the crucial observation that, in fact, {\it our  observable Universe is finite in size}\/, instead of extending over all of $\mathbb{R}^3$. The current value $R_0$ of the radius of the observable Universe provides us with a natural cutoff. In this way a stable vacuum state is guaranteed to exist.

\section{Newtonian cosmology as a quantum mechanics}

The Poisson equation satisfied by the nonrelativistic gravitational potential $U$ created by a mass density $\rho$,
\begin{equation}
\nabla^2U=4\pi G\rho,
\label{tretre}
\end{equation}
arises naturally in the weak--field limit of Einstein's field equations. In this limit, also called the {\it Newtonian approximation}\/, the (baryonic and dark) matter contents of the Universe is modelled as an ideal fluid (see, {\it e.g.}, the textbook \cite{WEINBERG}) satisfying the Poisson equation (\ref{tretre}) as well as the continuity equation  
\begin{equation}
\frac{\partial\rho}{\partial t}+\nabla\cdot\left(\rho{\bf v}\right)=0
\label{knoott}
\end{equation}
and the Euler equation
\begin{equation}
\frac{\partial{\bf v}}{\partial t}+\left({\bf v}\cdot\nabla\right){\bf v}+\frac{1}{\rho}\nabla p-{\bf F}=0.
\label{stella}
\end{equation}
In Eqs. (\ref{knoott}) and (\ref{stella}), $\rho$ is the volume density of fluid mass, $p$ is the pressure, ${\bf v}$ is the velocity field within the cosmological fluid, and ${\bf F}$ the force per unit volume acting on the fluid. The cosmological principle requires that the velocity ${\bf v}$ be everywhere proportional to the position vector ${\bf r}$. This latter statement is nothing but Hubble's law, which one can mimic by means of the phenomenological potential (\ref{potenzi}). Indeed the latter satisfies the Poisson equation (\ref{tretre}),
\begin{equation}
\nabla^2U_{\rm Hubble}=-3H_0^2,
\label{quattro}
\end{equation}
the right--hand side corresponding to a {\it negative}\/ mass density $\rho=-3mH_0^2/(4\pi G)$.

In ref. \cite{CABRERA} we have pointed out the existence of a remarkable {\it duality between nonrelativistic quantum mechanics on the one hand, and Newtonian cosmology on the other}\/ \cite{WIDROW}. Specifically, nonrelativistic quantum mechanics has a quantum probability fluid that exactly mimics the behaviour of the cosmological fluid, the latter considered in the Newtonian approximation. One proves that Eqs. (\ref{knoott}) and (\ref{stella}), which govern the cosmological fluid, become the very equations that govern the quantum probability fluid after applying the Madelung transformation. The inclusion of the Hubble potential as an external force acting on the quantum system then yields Eq. (\ref{tretre}).

The duality just mentioned can be used to {\it compute thermodynamical quantities of the Universe using standard quantum mechanics}\/.  In the introduction we have argued that the operator ${\bf R}^2=X^2+Y^2+Z^2$, which is proportional to the Hubble potential (\ref{potenzi}), is a measure of the amount of gravitational entropy enclosed by the Universe. Correcting dimensions by means of the appropriate physical constants, the operator
\begin{equation}
{\cal S}:={\cal N}\frac{k_B m H_0}{\hbar}{\bf R}^2
\label{nachhause}
\end{equation}
qualifies as a Boltzmann entropy. Above $m$ is the total mass (baryonic and dark) of the observable Universe. A {\it dimensionless}\/ factor ${\cal N}$ is left undetermined by these simple arguments; on general grounds we expect ${\cal N}$ to be of order unity. We call ${\cal S}$ the gravitational entropy operator.

The present paper is a continuation of, and an improvement on, our previous article \cite{CABRERA}. Let us examine this point in more detail. Within the scope of the approximations considered here, the effective Hamiltonian operator $H_{\rm eff}$ acting on the wavefunction $\psi({\bf r})$ that models the cosmological fluid is
\begin{equation}
H_{\rm eff}=-\frac{\hbar^2}{2m}\nabla^2-\frac{k_{\rm eff}}{2}{\bf r}^2, \qquad k_{\rm eff}=mH_0^2.
\label{jamilto}
\end{equation}
Above we have defined the effective elastic constant $k_{\rm eff}$ corresponding to the Hubble potential (\ref{potenzi}). The amount of mass $m_V$ contained within a volume $V$ equals $m_V=m\int_V{\rm d}^3x\vert\psi\vert^2$; the whole observable Universe is a sphere of radius $R_0$ (we collect our cosmological data $m$, $H_0$, $R_0$ from ref. \cite{PLANCK}).
Considering the Universe as a sphere with finite radius has the advantage that the instabilities \cite{BROADBRIDGE} due to the negative sign of the potential are avoided naturally. Although the Hamiltonian (\ref{jamilto}) can be diagonalised and its exact eigenfunctions can be obtained explicitly \cite{CABRERA, FINSTER}, the latter are extremely cumbersome for explicit computations. As a first step, for the sake of simplicity, in ref. \cite{CABRERA} we obtained the expectation value $\langle {\cal S}\rangle$ using a set of eigenfunctions of the {\it free}\/ Hamiltonian $-\hbar^2\nabla^2/(2m)$. 

The analysis performed in this paper uses the exact eigenfunctions of the effective Hamiltonian (\ref{jamilto}); this improves on the results of our calculation of ref. \cite{CABRERA}. The values thus obtained will be closer to actual (empirical) estimates for  the entropy of the Universe \cite{ASTROPH}, so the upper bound ${\cal S}_{\rm max}\sim 10^{123}k_B$ set by the holographic principle will no longer be saturated. Specifically, we will refine the results of our previous ref. \cite{CABRERA} by 3 orders of magnitude, see Eqs. (\ref{ergo}) and (\ref{kraft}) below.
Further work is required in order to extend our results beyond the Newtonian limit \cite{UPCOMING}; this extension will hopefully yield values in even better agreement with existing estimates.

\section{Estimate of the entropy}\label{npetrp}

Let us separate variables in the effective Hamiltonian (\ref{jamilto}) using spherical coordinates. The standard factorisation $\psi({\bf r})=R(r)Y_{lm}(\theta,\varphi)$ leads to a radial wave equation
\begin{equation}
\frac{1}{r^2}\frac{{\rm d}}{{\rm d}r}\left(r^2\frac{{\rm d}R}{{\rm d}r}\right)-\frac{l(l+1)}{r^2}R+\frac{2m}{\hbar^2}\left(E+\frac{k_{\rm eff}}{2}r^2\right)R=0.
\label{radas}
\end{equation}
The choice $l=0$ imposed by the cosmological principle leads to
\begin{equation}
r^2\frac{{\rm d}^2R}{{\rm d}r^2}+2r\frac{{\rm d}R}{{\rm d}r}+\frac{2m}{\hbar^2}\left(Er^2+\frac{mH_0^2}{2}r^4\right)R=0.
\label{vesel}
\end{equation}
As shown in refs. \cite{CABRERA, FINSTER}, two linearly independent solutions of (\ref{vesel}) turn out to be
\begin{equation} 
R_{\alpha}^{(1)}(r)=\exp\left(\frac{{\rm i}\beta^2r^2}{2}\right) {}_1F_1\left(\frac{3}{4}-\frac{{\rm i}\alpha}{4},\frac{3}{2}; -{\rm i}\beta^2r^2\right)
\label{herri}
\end{equation}
and
\begin{equation}
R_{\alpha}^{(2)}(r)=\frac{1}{r}\exp\left(\frac{{\rm i}\beta^2r^2}{2}\right){}_1F_1\left(\frac{1}{4}-\frac{{\rm i}\alpha}{4},\frac{1}{2}; -{\rm i}\beta^2r^2\right),
\label{tabernae}
\end{equation}
where ${}_1F_1(a,b;z)$ is the confluent hypergeometric function \cite{LEBEDEV}, and the parameters $\alpha$, $\beta$ take on the values
\begin{equation}
\alpha:=\frac{2E}{\hbar H_0},\qquad \beta^4:=\frac{m^2H_0^2}{\hbar^2}.
\label{morcilla}
\end{equation}

To begin with, the complete wavefunction corresponding to the radial wavefunction (\ref{herri}) reads
\begin{equation}
\psi_{\alpha}^{(1)}(r,\theta,\varphi)=\frac{N_{\alpha}^{(1)}}{\sqrt{4\pi}}\exp\left(\frac{{\rm i}\beta^2r^2}{2}\right) {}_1F_1\left(\frac{3}{4}-\frac{{\rm i}\alpha}{4},\frac{3}{2}; -{\rm i}\beta^2r^2\right);
\label{karkon}
\end{equation}
the radial normalisation factor $N_{\alpha}^{(1)}$ will be determined presently. The eigenfunction $\psi_{\alpha}^{(1)}$ is singularity free over the entire interval $[0,R_0]$. A numerical estimate yields $\beta\simeq 1.1\times 10^{35}$ metres${}^{-1}$. Given that $R_0\simeq 4.4\times 10^{26}$ metres, the dimensionless product $(\beta r)^2$ in Eq. (\ref{karkon}) quickly drives the function ${}_1F_1$ into its asymptotic regime, where it can be approximated as \cite{LEBEDEV}
\begin{equation}
{}_1F_1(a,b;z)\simeq\frac{\Gamma(b)}{\Gamma(b-a)}{\rm e}^{-{\rm i}\pi a}z^{-a}
+\frac{\Gamma(b)}{\Gamma(a)}{\rm e}^z\,z^{a-b},\quad \vert z\vert\to\infty,
\label{toninfest}
\end{equation}
whenever $\vert{\rm arg}(z)\vert<\pi$ and $b\neq 0,-1,-2,\ldots$ We will also need Stirling's formula 
\begin{equation}
\Gamma(t)\simeq\exp\left[\left(t-\frac{1}{2}\right)\ln t -t+\frac{1}{2}\ln 2\pi\right],
\label{gamagrande}
\end{equation}
valid for $\vert t\vert\to\infty$ whenever $\vert{\rm arg}(t)\vert<\pi$. When applying Stirling's approximation we will select the main branch of the complex logarithm. Another order--of--magnitude estimate yields $\alpha\simeq 10^{52} E$, with the energy $E$ expressed in Joule; this fact allows to drop the first summand in (\ref{toninfest}) in favour of the second. Then a  lengthy calculation based on Eqs. (\ref{toninfest}) and (\ref{gamagrande}) yields the desired asymptotic expression of the confluent hypergeometric function in (\ref{karkon}):
$$
{}_1F_1\left(\frac{3}{4}-\frac{{\rm i}\alpha}{4},\frac{3}{2}; -{\rm i}\beta^2r^2\right)\simeq\frac{1}{2\sqrt{2}}\exp\left(\frac{3}{4}-{\rm i}\pi\right)
\exp\left(\frac{\pi\alpha}{2}\right)\exp\left(\frac{{\rm i}\alpha}{4}\ln\frac{\alpha}{4}\right)
$$
\begin{equation}
\times\exp\left\{-{\rm i}\left[\beta^2 r^2+\frac{\alpha}{2}\ln (\beta r) \right]\right\}\exp\left(-\frac{3}{2}\ln \beta r\right),\qquad r\to\infty.
\label{tertia}
\end{equation}
{}Finally substituting Eq. (\ref{tertia}) into Eq. (\ref{karkon}), and absorbing an irrelevant constant within the normalisation factor $N_{\alpha}^{(1)}$, we obtain the following asymptotic wavefunction:
$$
\psi_{\alpha}^{(1)}(r,\theta,\varphi)\simeq\frac{N_{\alpha}^{(1)}}{\sqrt{4\pi}}
\exp\left(\frac{\pi\alpha}{2}\right)
\exp\left(\frac{{\rm i}\alpha}{4}\ln\frac{\alpha}{4}\right)
$$
\begin{equation}
\times
\exp\left\{-\frac{{\rm i}}{2}\left[\alpha\ln(\beta r) +\beta^2r^2\right]\right\}(\beta r)^{-3/2},\qquad r\to\infty.
\label{quarta}
\end{equation}
We observe that the asymptotic expression (\ref{quarta}) is singular at $r=0$ while the original wavefunction (\ref{karkon}) was not. This is just a consequence of having replaced the exact wavefunction with its asymptotic approximation for large $r$. Therefore Eq. (\ref{quarta}) applies at most over the interval $[\epsilon, R_0]$, where $\epsilon>0$ is small but nonvanishing. We need to determine a suitable $\epsilon$ and the wavefunction $\psi_{\alpha}^{(1)}$ over $[0,\epsilon]$. 

A natural choice to make is $\epsilon=\beta^{-1}$. This is sufficiently small while, at the same time, values of $r>\beta^{-1}$ fall well within the asymptotic regime (\ref{toninfest}) of the confluent hypergeometric function. Over the interval $[0,\beta^{-1}]$ we will approximate ${}_1F_1$ by its Taylor expansion ${}_1F_1(a,b;z)\simeq 1+az/b$ \cite{LEBEDEV}. Altogether the normalised, approximate wavefunction for the matter contents of the Universe 
$$
\psi_{\alpha}^{(1)}(r,\theta,\varphi)=\sqrt{\frac{\beta^3}{4\pi\ln \left(\beta R_0\right)}}
\exp\left(\frac{{\rm i}\alpha}{4}\ln\frac{\alpha}{4}\right)
$$
\begin{equation}
\times\left\{\begin{array}{ll}
\exp\left(-{\rm i}/{2}\right),\quad\qquad\quad\quad\qquad\qquad\qquad\qquad\;\; r\in[0,\beta^{-1}]\\
\exp\left\{-\frac{{\rm i}}{2}\left[\alpha\ln (\beta r) +\beta^2r^2\right]\right\}
\left(\beta r\right)^{-3/2},\qquad\;  r\in[\beta^{-1},R_0]
\end{array}\right.
\label{octavia}
\end{equation}
is regular over the entire interval $[0,R_0]$. With the wavefunction (\ref{octavia}) we obtain
\begin{equation}
\langle\psi_{\alpha}^{(1)}\vert {\bf R}^2\vert\psi_{\alpha}^{(1)}\rangle=\frac{R_0^2}{2\ln \left(\beta R_0\right)},
\label{incontro}
\end{equation}
after dropping subleading terms in $\beta$. Substituted back into Eq. (\ref{nachhause}), this produces a value of the entropy
\begin{equation}
\langle\psi_{\alpha}^{(1)}\vert {\cal S}\vert\psi_{\alpha}^{(1)}\rangle=6{\cal N}\times10^{120} k_B
\label{ergo}
\end{equation}
which, taking ${\cal N}=1/6$, is three orders of magnitude below the upper bound ${\cal S}_{\rm max}\sim10^{123}k_B$ set by the holographic principle. This is a considerable improvement on the results of ref. \cite{CABRERA}, where the holographic bound  was saturated.

In the case of the second, linearly independent radial wavefunction (\ref{tabernae}) we have the complete eigenfunction
\begin{equation}
\psi_{\alpha}^{(2)}(r,\theta,\varphi)=\frac{N_{\alpha}^{(2)}}{\sqrt{4\pi}}\frac{1}{r}\exp\left(\frac{{\rm i}\beta^2r^2}{2}\right){}_1F_1\left(\frac{1}{4}-\frac{{\rm i}\alpha}{4},\frac{1}{2}; -{\rm i}\beta^2r^2\right).
\label{oeko}
\end{equation}
As opposed to $\psi_{\alpha}^{(1)}$, the wavefunction $\psi_{\alpha}^{(2)}$ is singular at $r=0$. Again applying Eqs. (\ref{toninfest}) and (\ref{gamagrande}) one finds the asymptotics
\begin{equation}
{}_1F_1\left(\frac{1}{4}-\frac{{\rm i}\alpha}{4},\frac{1}{2};-{\rm i}\beta^2r^2\right)\simeq\frac{1}{\sqrt{2}}\exp\left(\frac{1}{4}-\frac{{\rm i}\pi}{2}\right)\exp\left(\frac{\pi\alpha}{2}+\frac{{\rm i}\alpha}{4}\ln\frac{\alpha}{4}\right)
\label{dosefe}
\end{equation}
$$
\times\exp\left[-{\rm i}\left(\frac{\alpha}{2}\ln\beta r + \beta^2 r^2\right)\right]\exp\left(-\frac{1}{2}\ln \beta r\right),\quad r\to\infty.
$$
Next substituting (\ref{dosefe}) into (\ref{oeko}) produces, after absorbing an irrelevant constant within the normalisation factor,
\begin{equation}
\psi_{\alpha}^{(2)}(r,\theta,\varphi)\simeq\frac{N_{\alpha}^{(2)}}{\sqrt{4\pi}}\frac{1}{r}
\exp\left(\frac{\pi\alpha}{2}+\frac{{\rm i}\alpha}{4}\ln\frac{\alpha}{4}\right)
\label{hellas}
\end{equation}
$$
\times\exp\left[-\frac{{\rm i}}{2}\left(\alpha\ln\beta r+\beta^2r^2\right)\right]
(\beta r)^{-1/2},\quad r\to\infty.
$$
{}Finally, arguments similar to those leading up to Eq. (\ref{octavia}) produce the following normalised, approximate wavefunction over the complete interval $[0, R_0]$:
$$
\psi_{\alpha}^{(2)}(r,\theta,\varphi)=\sqrt{\frac{\beta}{4\pi\ln(\beta R_0)}}\exp\left(\frac{{\rm i}\alpha}{4}\ln\frac{\alpha}{4}\right)
$$
\begin{equation}
\times\left\{\begin{array}{ll}
\frac{1}{r}\exp\left(-{\rm i}/{2}\right),\quad\quad\quad\qquad\qquad\qquad\qquad\qquad\;\; r\in[0,\beta^{-1}]\\
\frac{1}{r}\exp\left\{-\frac{{\rm i}}{2}\left[\alpha\ln(\beta r) +\beta^2r^2\right]\right\}
\left(\beta r\right)^{-1/2},\qquad\;  r\in[\beta^{-1},R_0].
\end{array}\right.
\label{tredicesima}
\end{equation}
We observe that the approximate wavefunction (\ref{tredicesima}) remains singular at $r=0$, as imposed by the exact wavefunction (\ref{oeko}). With the above one computes
\begin{equation}
\langle\psi_{\alpha}^{(2)}\vert{\bf R}^2\vert\psi_{\alpha}^{(2)}\rangle=\frac{R_0^2}{2\ln(\beta R_0)},
\label{lola}
\end{equation}
coincident with the corresponding result (\ref{incontro}) for the regular wavefunction. Therefore
\begin{equation}
\langle\psi_{\alpha}^{(2)}\vert {\cal S}\vert\psi_{\alpha}^{(2)}\rangle=6{\cal N}\times10^{120} k_B,
\label{kraft}
\end{equation}
in complete agreement with the entropy already found in (\ref{ergo}) for the regular wavefunction.

\section{Discussion}

The holographic principle sets an upper bound of approximately $10^{123}k_B$ on the entropy content of the Universe. Some phenomenological estimates \cite{ASTROPH} place the actual value at around $10^{104}k_B$, gravitational entropy (and, in particular, black holes) representing the largest single contributors to the entropy budget of the Universe. Although Newtonian cosmology does allow for black holes \cite{MEXICO}, the many simplifications made by our elementary model necessarily leave out some essential physics of the Universe. Nevertheless, our toy model succeeds in capturing some key elements of reality. For example,  the upper bound set by the holographic principle is always respected, even by such a crude approximation as the free waves \cite{CABRERA}. The Hubble waves (\ref{octavia}) and (\ref{tredicesima}) represent a considerable improvement on the free waves, as they reduce the expectation value of the entropy by three orders of magnitude. We hope that a fully general--relativistic treatment \cite{UPCOMING} will yield results in even better agreement with existing empirical estimates. 

Admittedly, solutions (\ref{herri}) and (\ref{tabernae}) violate the cosmological principle. In fact any solution to the (interacting) Schroedinger equation will violate the cosmological principle. Only free wave solutions to the free wave equation ({\it i.e.}\/, with zero potential) satisfy the cosmological principle. However, the free wavefunctions of our previous ref. \cite{CABRERA} saturate the holographic principle, while our improved Hubble wavefunctions (\ref{herri}) and (\ref{tabernae}) no longer saturate it. This is essential for the very existence of life in the Universe. Given that the cosmological principle itself is an idealisation, we believe the improved entropy results obtained using Hubble wavefunctions outweigh the violation of the cosmological principle.

Since $\alpha$ in Eq. (\ref{morcilla}) is the (dimensionless) energy eigenvalue in $H_{\rm eff}\psi=E\psi$, the parameter $\alpha$ plays the same role that the quantum number $n\in\mathbb{N}$ plays in the standard harmonic oscillator, where the potential energy is positive definite. Our negative definite harmonic potential does not have quantised energy levels, but continuous energy levels $\alpha$ instead. However the range of values covered by $\alpha$, while unbounded above, is bounded below by the existence of the radius of the Universe: a classical particle at rest at $r=R_0$ would carry an energy
\begin{equation}
E_0=-\frac{1}{2}mH_0^2R_0^2.
\label{zpetite}
\end{equation}
This configuration can be regarded as the classical vacuum state. In terms of the dimensionless eigenvalue $\alpha$, this energy equals
\begin{equation}
\alpha_0=-\frac{mH_0R_0^2}{\hbar}=-2.6\times 10^{123}.
\label{zteddy}
\end{equation}
The vacuum energy (\ref{zteddy}) has been determined by a classical argument; although the uncertainty principle will shift the minimum energy (\ref{zteddy}) by a positive amount, this correction can be discarded for our purposes, as it will be negligible compared to (\ref{zteddy}) itself. The negative sign in (\ref{zteddy}) is due to the Hubble potential (\ref{potenzi}), while the dimensionless factor $2.6$ is of order unity. Thus the vacuum energy (\ref{zteddy}) yields the approximate equality
\begin{equation}
\vert\alpha_0\vert\simeq\frac{{\cal S}_{\rm max}}{k_B}\simeq10^{123}.
\label{buzle}
\end{equation}
The above numerical coincidence is in fact a consistency check on all our previous arguments. It confirms once again that the holographic bound never gets exceeded, since both the energy and the entropy grow quadratically with the distance.

We have seen in section \ref{npetrp} that the linearly independent wavefunctions $\psi_{\alpha}^{(1)}$ and $\psi_{\alpha}^{(2)}$ 
coalesce asymptotically in $r$. This occurs despite the fact that $\psi_{\alpha}^{(1)}$ is regular at $r=0$ while $\psi_{\alpha}^{(2)}$ is singular. In turn, this implies that issues of regularity of the wavefunction at $r=0$ are irrelevant for our purposes. Our estimate of the entropy remains valid regardless of the precise wavefunction used in a neighbourhood of $r=0$; this neighbourhood is $[0,\beta^{-1}]$.

The constant $\beta$ arises naturally when diagonalising the effective Hubble Hamiltonian (\ref{jamilto}), see Eq. (\ref{morcilla}). It turns out that $\beta^{-1}\simeq10^{-35}$ metres, which is close to the value of the Planck length $L_P$,
\begin{equation}
\beta^{-1}=\sqrt{\frac{\hbar}{mH_0}}\simeq L_P=\sqrt{\frac{\hbar G}{c^3}}.
\label{apross}
\end{equation}
Our toy model of the Universe thus possesses an intrinsic length scale, $\beta^{-1}$, which numerically equals the Planck length. This approximate equality is no coincidence: the value of $m$ is that of the mass enclosed by the Hubble horizon for a critical Universe, $m\simeq 1/(H_0G)$, hence $\beta\simeq 1/\sqrt{G}=1/L_P$.

Our analysis is rooted in previous studies  \cite{ELZE,GALLEGO} on the emergent property of quantum mechanics. According to the hypothesis of emergence, quantum mechanics as we know it should be the effective theory of some underlying mechanics, the coarse graining of which would yield our current quantum models. Important recent work in general relativity \cite{PADDY1,PADDY2,PADDY3,PADDY4} also points in the same direction: gravity appears to be the {\it thermodynamics}\/ of some underlying degrees of freedom, a continuous spacetime emerging only as their low--energy limit. That seemingly unrelated fields such as quantum theory and general relativity might share fundamental common features \cite{MATONEGRAVITY} is an intriguing possibility worthy of future study.

\vskip0.5cm
\noindent
{\bf Acknowledgements}  This research was supported by grant no. ENE2015-71333-R (Spain).

\end{document}